\documentclass[usenatbib,usegraphicx]{mn2e}
\usepackage{subfigure}
\usepackage{longtable}
\usepackage{graphicx}
\usepackage{graphics}
\usepackage{keyval}
\usepackage{trig}
\usepackage{rotating}
\def\beq{\begin{equation}}
\def\eeq{\end{equation}}
\def\bey{\begin{eqnarray}}
\def\eey{\end{eqnarray}}

\def\msun{M_\odot}
\def\sun{\odot}
\def\lsim{\mathrel{\raise.3ex\hbox{$<$\kern-.75em\lower1ex\hbox{$\sim$}}}}
\def\gsim{\mathrel{\raise.3ex\hbox{$  $\kern-.75em\lower1ex\hbox{$\sim$}}}}

\def\lsun{L_\odot}
\def\kms{\, {\rm km \, s}^{-1} }

\title{Dwarf Spheroidals in MOND}
\author[G. W. Angus]{G. W. Angus$^{1}$\thanks{email:
gwa2@st-andrews.ac.uk}   \\
$^{1}$SUPA, School of Physics and Astronomy, University of St. Andrews, KY16 9SS Scotland\\ }

\begin{document}

\date{Accepted ... Received ... ; in original form ...}

\pagerange{\pageref{firstpage}--\pageref{lastpage}} \pubyear{2007}

\maketitle

\label{firstpage}
\begin{abstract}
We take the line of sight velocity dispersions as functions of radius for 8 Milky Way dwarf spheroidal galaxies and use Jeans analysis to calculate the mass-to-light ratios (M/L) in Modified Newtonian Dynamics (MOND). Using the latest structural parameters, distances and variable velocity anisotropy, we find 6/8 dwarfs have sensible M/L using only the stellar populations. Sextans and Draco, however, have M/L=$9.2_{-3.0}^{+5.3}$ and $43.9_{-19.3}^{+29.0}$ respectively, which poses a problem. Apart from the need for Sextans' integrated magnitude to be reviewed, we propose tidal effects intrinsic to MOND, testable with numerical simulations, but fully orbit dependant,  which are disrupting Draco. The creation of the Magellanic Stream is also re-addressed in MOND, the scenario being the stream is ram pressure stripped from the SMC as it crosses the LMC.
\end{abstract}


\section{Introduction}
\protect\label{sec:intr}
In addition to the obvious Magellanic Clouds and the tidally disrupting Sagittarius dwarf (Ibata, Gilmore \& Irwin 1994), the Milky Way (MW) has 8 relatively nearby dwarf spheroidal satellite galaxies for which the radial velocities of many hundreds of member stars have been meticulously measured at various projected radii allowing the calculation of the line of sight (los) velocity dispersion as a function of projected radius. If we assume the satellites are spherical and in equilibrium, we can use specific mass models to check the consistency of the luminous matter in replicating the los dispersion profiles through the Jeans equation. 

It so happens that these small satellites appear to be some of the most heavily cold dark matter (CDM) dominated systems in the universe and so including only the luminous matter in Jeans analysis underestimates the los velocity dispersions considerably. It is because of the well studied, small spacial scales and large dark matter content that Gilmore et al. (2007ab) have used the relatively low dark matter density in the cores of dwarf spheroidals to argue for an upper limit on the mass of the DM particles from supersymmetric theories.

There has been speculation that Ursa Minor and Draco are formed from the tidal debris of the Magellanic Stream (Lynden-Bell 1983; Metz \& Kroupa 2007) and that Leo I, Leo II and Sculptor have similarly formed from debris cast by the tidal break-up of a previously larger Fornax satellite. More likely, the dwarfs were part of two small groups of galaxies that have entered the Milky Way at late times.

Dwarf galaxies often form from the debris of tidally disrupted progenitor galaxies. In this case, very little CDM accompanies the stars and gas for the simple reason that the tidal disruption must have torn the gas and the CDM out into streams from which the gas has condensed and formed stars and only a limited amount of CDM can have condensed with it. This is contrary to the standard picture where gas cools and falls onto an already formed CDM halo.

There has been a recent study by Bournaud et al. (2007) of tidal dwarf galaxies formed from the debris of NGC5291. They found a persisting mass discrepancy by measuring the visible galactic masses and comparing with the mass implied by the rotation curves. They proposed the existence of hitherto unheard of disk DM, in the progenitor galaxies, in the form of molecular gas (as opposed to the cold DM of massive particle form). Two simultaneous studies (Milgrom 2007 and Gentile et al. 2007) in the framework of Milgrom (1983)'s Modified Newtonian Dynamics (MOND: for reviews see Sanders \& McGaugh 2002; Bekenstein 2006; Milgrom 2008) found the discrepancy was exactly that expected in MOND because the tidal dwarfs are very diffuse and low surface brightness (McGaugh \& de Blok 1998) meaning they have low internal accelerations (much less than the transition acceleration of the theory $a_o = 1.2\times10^{-10}ms^{-2}=3.6(\kms)^2 pc^{-1}$. In addition, the external accelerations from the large progenitors is also weak (discussed in depth in both papers) allowing a relatively large boost to the internal gravity enabling perfect agreement with the observed rotation curve using just the luminous matter (no need for CDM or molecular disk DM). This is the most remarkable evidence for MOND since the baryonic Tully-Fisher relation was matched in great detail (McGaugh et al. 2000; McGaugh 2005).

Previous attempts to check the consistency of the Milky Way's dwarfs with MOND have been limited to single number central velocity dispersions (although see Lokas et al. 2001 which sadly does not include the external field effect). This gave good estimates of the mass-to-light ratios (M/L) for all the dwarfs studied here by considering whether each dwarf was dominated gravitationally by the external field of the Milky Way (MW) or by its own internal gravity (see Milgrom 1995, Brada \& Milgrom 2000). If it was dominated by the external gravity, $g_{ex}$ then the Newtonian M/L was used after renormalisation of Newton's constant, $G_{eff}/G = a_o/g_{ex}$, reducing the M/L in MOND by the same factor. If the external gravity was weak w.r.t. the internal gravity then the dwarf mass was calculated from the relation $M={81 \over 4}{\sigma_{los}^4 \over Ga_o}$ from which the M/L can be deduced. Nevertheless, proper Jeans modelling with the luminosity profile of the dwarf along with the near exact calculation of the dwarf's internal gravity at all radii by including analysis of the external field effect (see Milgrom 1983, Angus \& McGaugh 2008ab and Wu et al. 2007) and variation of the velocity anisotropy profile (Angus et al. 2008) in order to match the observed los dispersions at all radii is far more satisfactory. This also gives tighter limits on the M/L.

\section{Modeling the los velocity dispersions}
The key data for the analysis are the luminosity of the dwarf, its structural profile and the Galactocentric distance. The total luminosities, $L_v$ have been measured only in the V-band and some, especially that of Sextans, are very old and we suggest that new K-band images would be more appropriate measures of the stellar mass (Driver et al. 2007).

Next we use the King models (concentration c and tidal radius $R_t$) for the enclosed luminosity profiles, $L_K(r)$, of the dwarfs (mostly taken from Irwin \& Hatzidimitriou 1995) that allows us to numerically calculate the luminosity density as a function of radius. Finally, we need the current Galactocentric radius, $R_{MW}$ in order to compute the external gravity from the MW via the flat rotation curve (see Angus \& McGaugh 2008b) $g_{ex}={(170\kms)^2 \over R_{MW}}$ where the amplitude of the rotation curve is from the SDSS data compiled by Xue et al. (2008). From this we solve for the internal gravity
\beq
\protect\label{eqn:mond}
g(r) \mu \left(|g(r)+g_{ex}|/a_o\right)=G(M/L)L_K(r)r^{-2}
\eeq
where we use the $\mu$ function employed by Famaey \& Binney (2005) in their fit to the MW's terminal velocity curve
\beq
\protect\label{eqn:mu}
\mu(x)=x/(1+x),
\eeq
and we take $|g(r)+g_{ex}|=\sqrt{g(r)^2+g_{ex}^2}$ which only matters in the small transition region. Then we numerically differentiate the luminosity profile from the King model, $L_K(r)$, to match with the simple function 
\beq
\protect\label{eqn:lum}
l(r) \propto \left(1+{r \over r_{\alpha}}\right)^{-\alpha_o}
\eeq
which is a luminosity density profile and helps us to analytically define $\alpha(r)=d\ln l(r) / d \ln r=-\alpha_o{r \over r+r_{\alpha}}$ for solving Jeans equation for the radial velocity dispersions, $\sigma_r$,
\beq
\protect\label{eqn:jeans}
{d \over dr}\sigma_r^2(r) + {\alpha(r)+2\beta(r) \over r}\sigma_r^2(r) = -g(r)
\eeq
and we can cast this into the los with the use of the luminosity density profile, $l(r)$ (see Binney \& Mamon 1982) as per Angus et al. (2008). Taking an anisotropy, $\beta$ of the form
\beq
\protect\label{eqn:beta}
\beta(r)=\beta_o-{r \over r_{\beta}}.
\eeq
 This and all other fit parameters are given in table 1.

Here we have three free parameters available for matching the observed and model velocity dispersions. They are the M/L and the two associated with $\beta$. M/L adjusts the normalisation, whereas $\beta$ varies the shape of the los profile. Note that although there are two parameters for the anisotropy, the profile required is virtually the same for 6/8 dwarfs (see fig 2). In any case, Koch et al. (2007) make it clear from their analysis of Leo II that there is compelling evidence for velocity anisotropy in dwarfs.

\begin{figure*}
\def\subfigtopskip{0pt} 
\def\subfigbottomskip{4pt}
\def\subfigcapskip{1pt}
\centering

\begin{tabular}{ccc}

\subfigure{\label{fig:carina}
\includegraphics[angle=0,width=6.0cm]{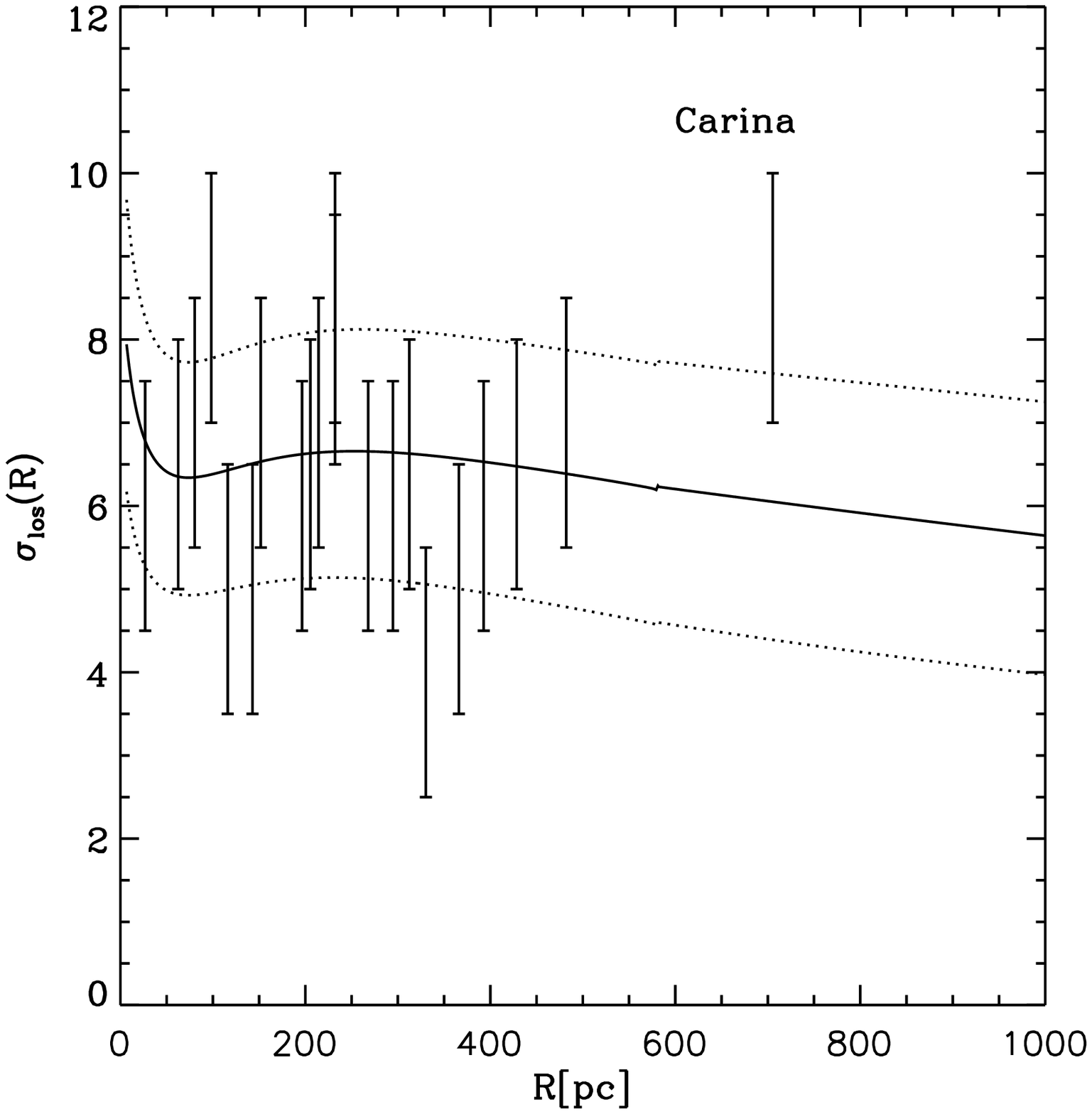}
}
&
\subfigure{\label{fig:draco}
\includegraphics[angle=0,width=6.0cm]{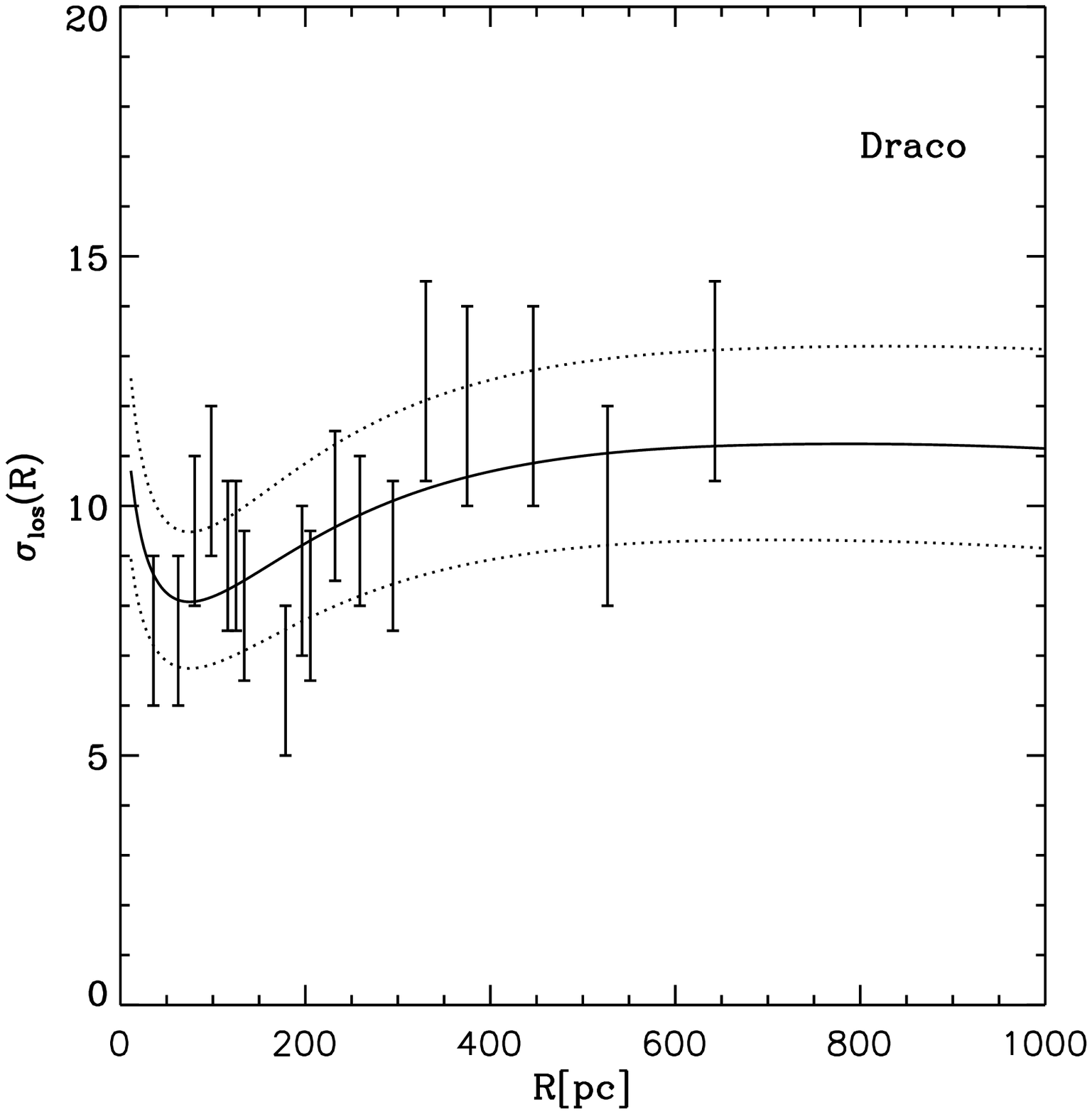}
}
&
\subfigure{\label{fig:leo1}
\includegraphics[angle=0,width=6.0cm]{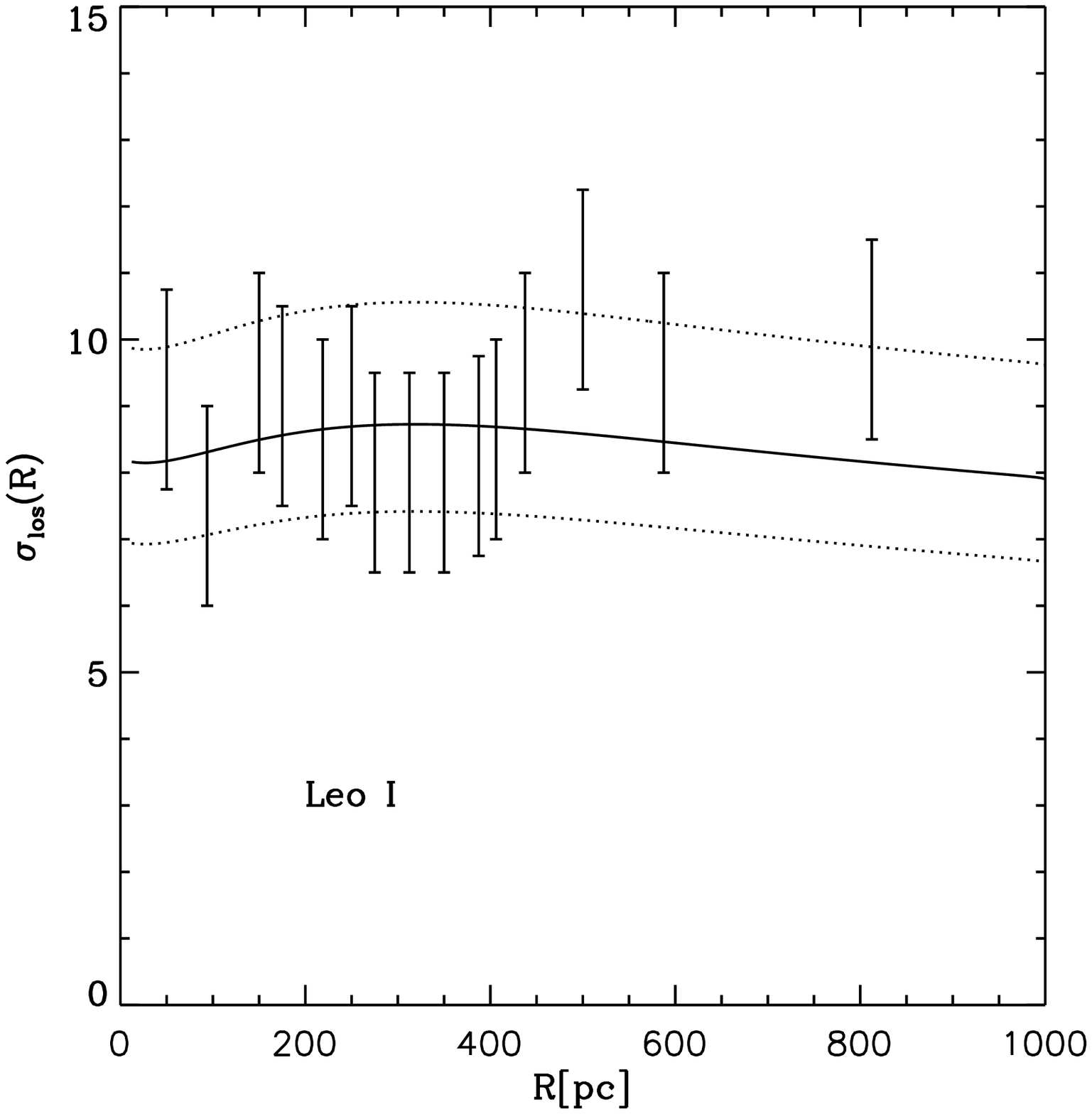}
}
\\
\subfigure{\label{fig:sextans}
\includegraphics[angle=0,width=6.0cm]{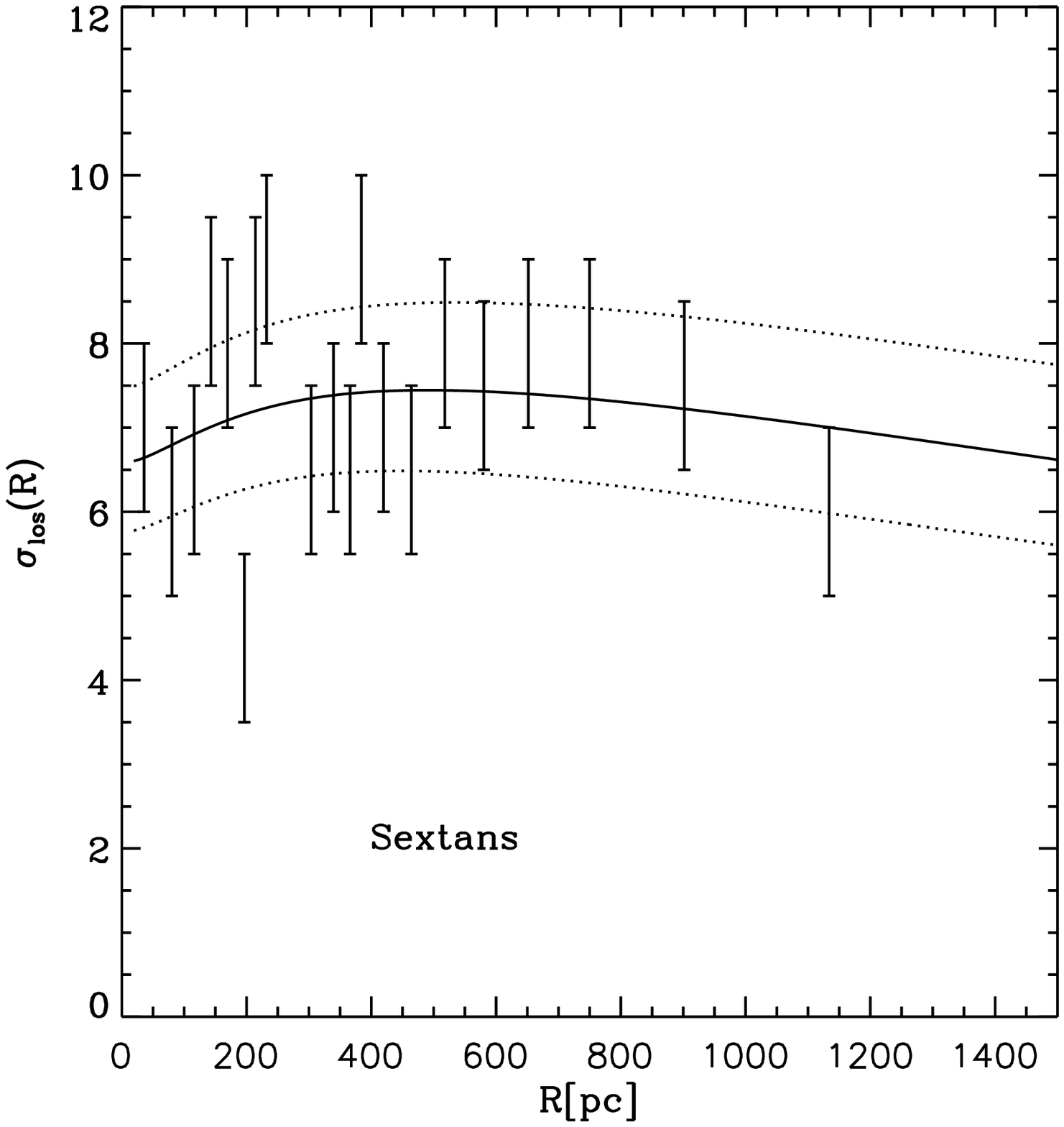}
}
&
\subfigure{\label{fig:fornax}
\includegraphics[angle=0,width=6.0cm]{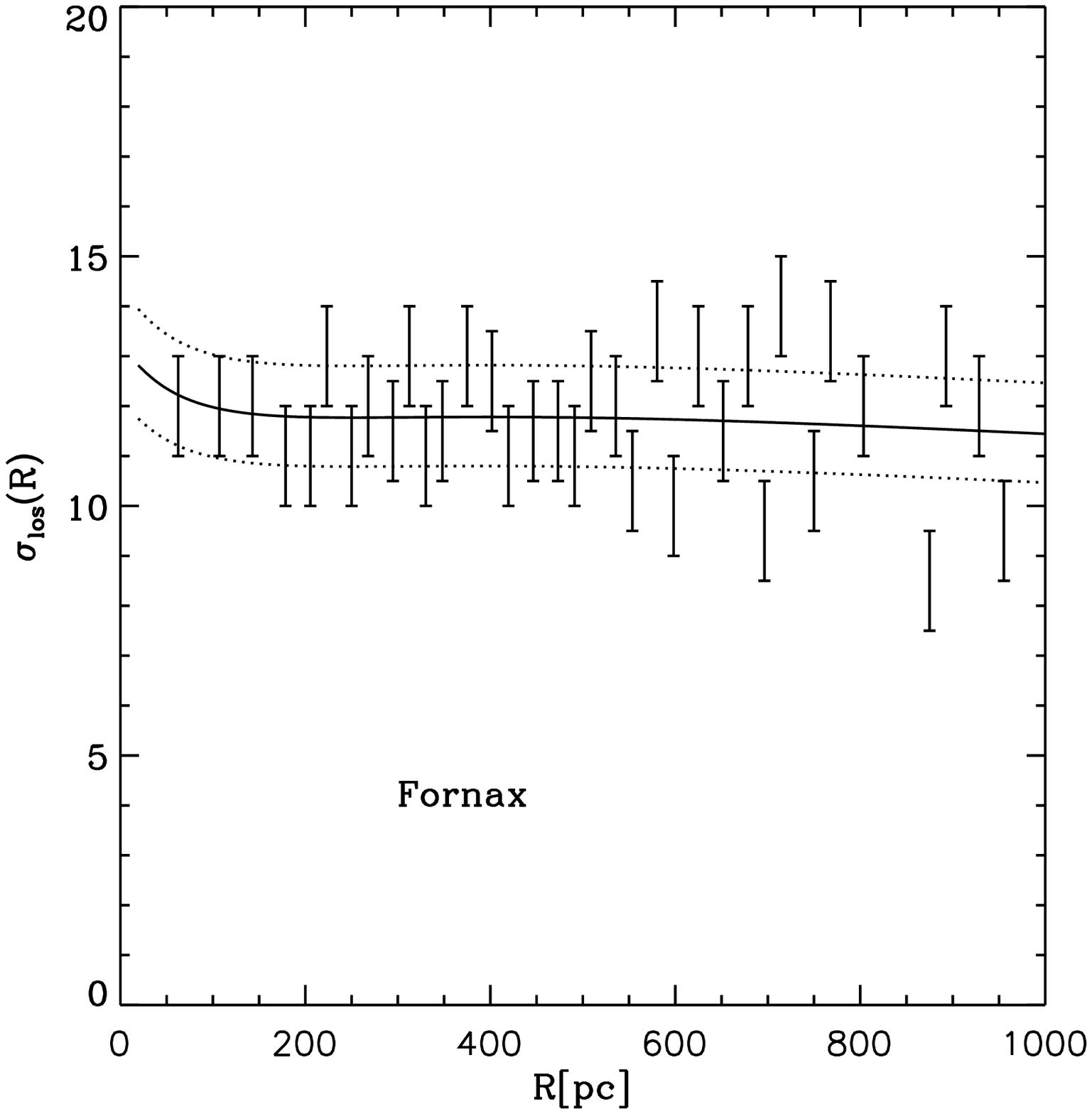}
}
&
\subfigure{\label{fig:sculptor}
\includegraphics[angle=0,width=6.0cm]{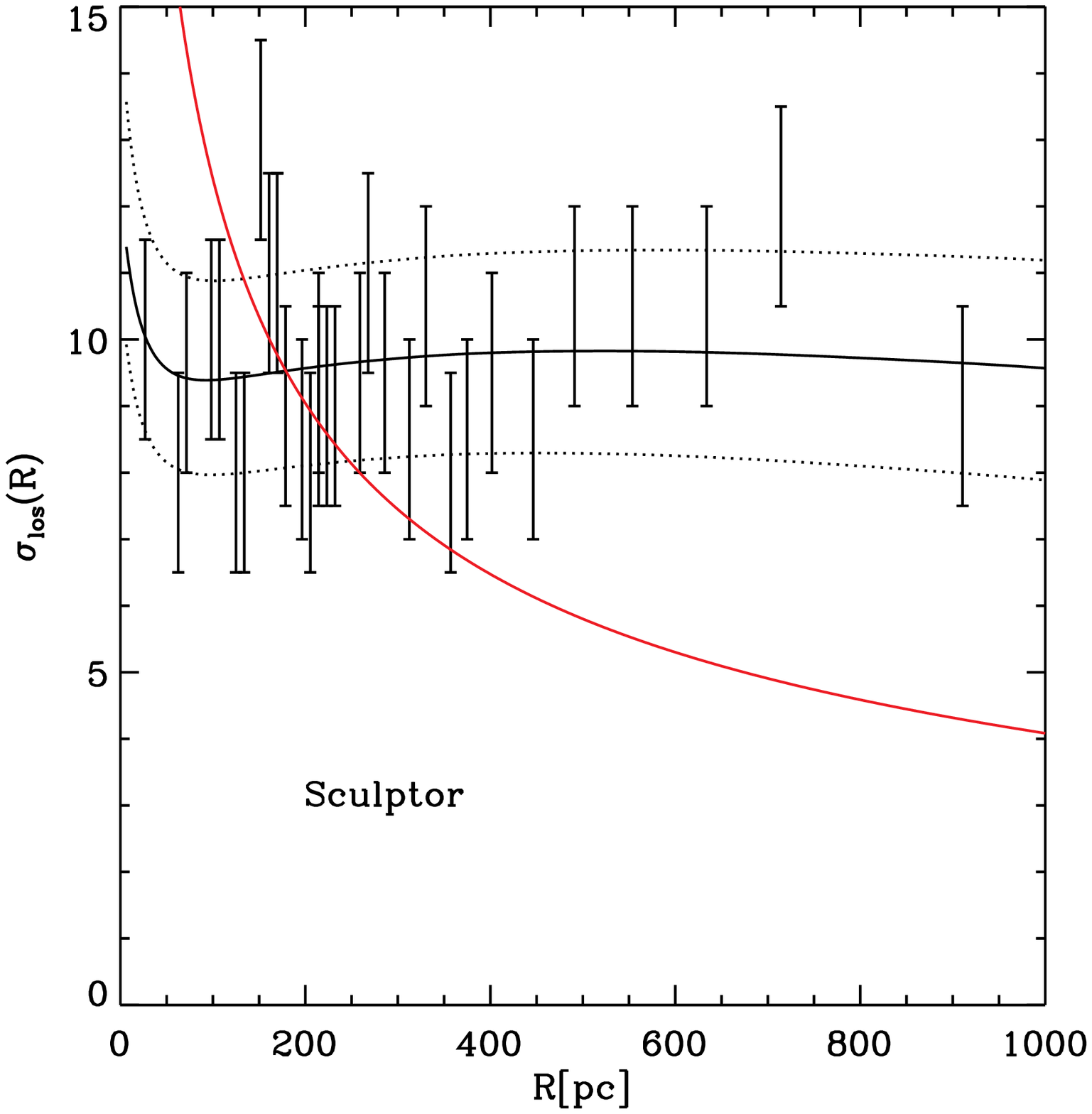}
}\\
\subfigure{\label{fig:leo2}
\includegraphics[angle=0,width=6.0cm]{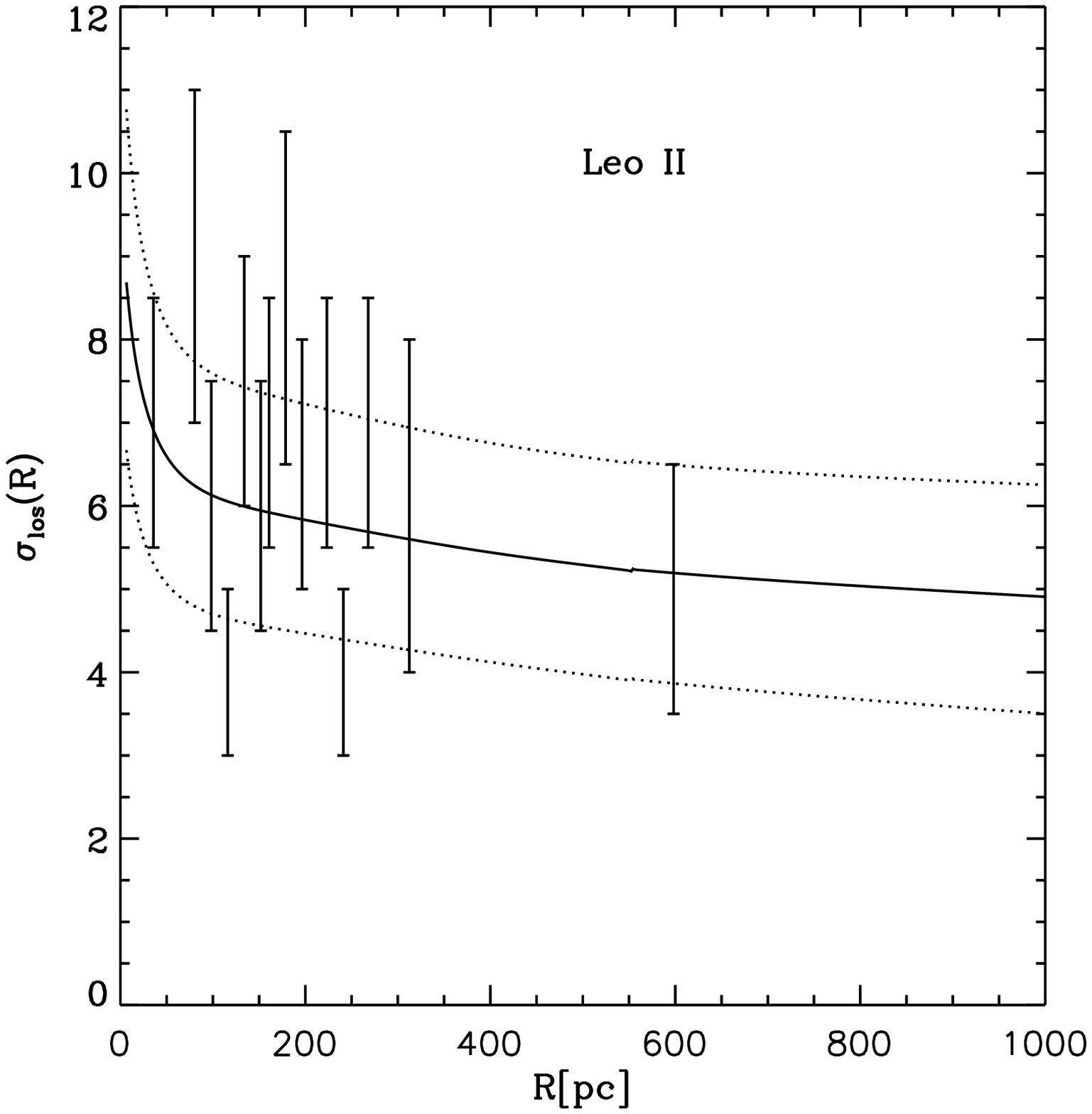}
}
&
\subfigure{\label{fig:umi}
\includegraphics[angle=0,width=6.0cm]{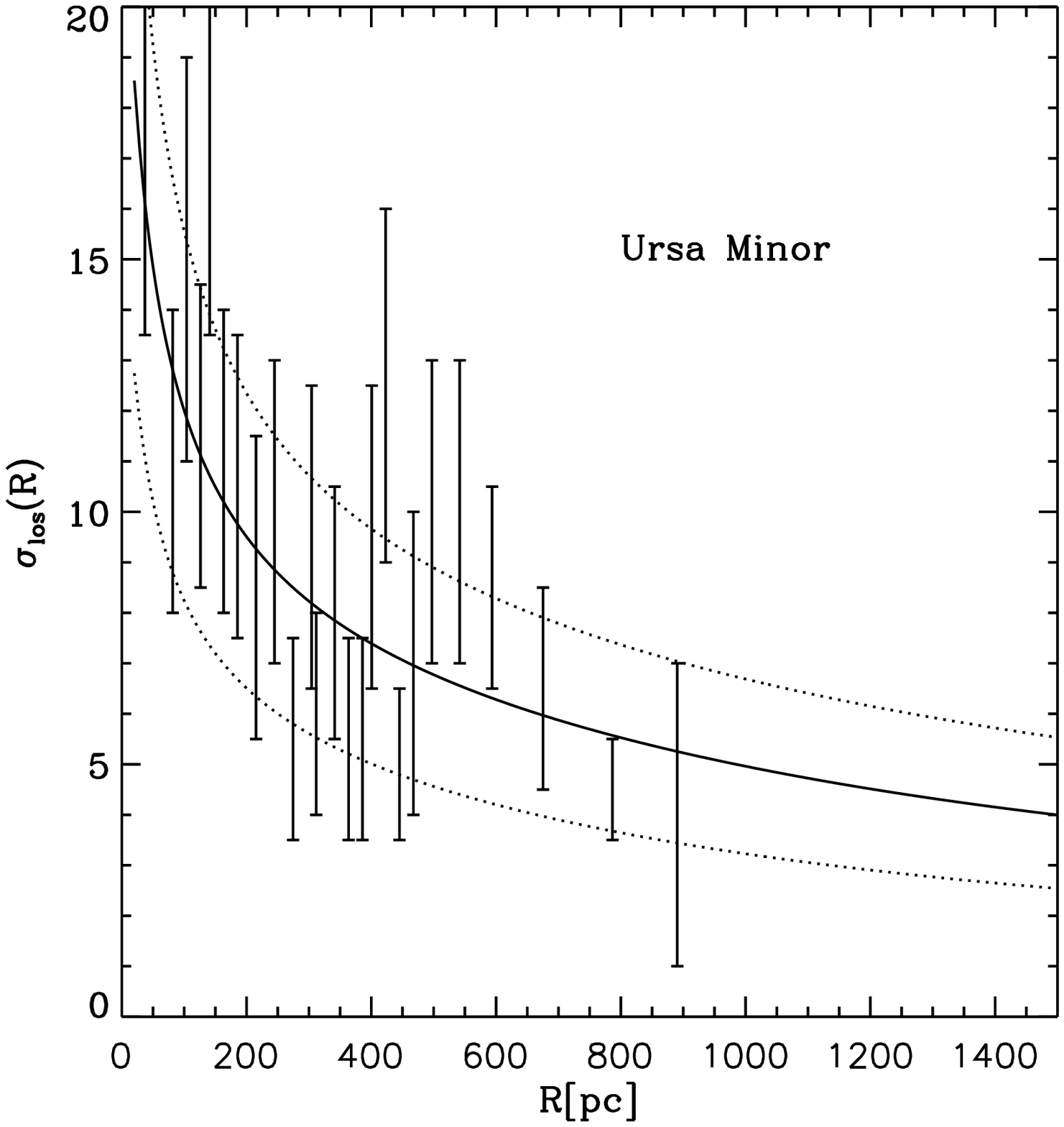}
}\\
\end{tabular}

\caption{Velocity dispersion profiles, $\sigma_{los}$(R), for all 8 dwarfs. Best fit lines have solid line type and the dashed curves correspond to the 1-sigma errors on M/L.}
\label{fig:dsph}
\end{figure*}

By $\chi^2$ fitting the los velocity dispersions of Mateo, Olszewski \& Walker (2007) for Leo I, Mu\~noz et al. (2005) for Ursa Minor and Walker et al. (2007) for the remaining 6, with the model MOND dispersions, we were able to best fit the anisotropy profile of all 8 dwarfs and thereby compute the M/L. The fits are not hugely sensitive to the anisotropy profile. For instance, in table 1 the column of $r_{\beta}$ also gives the 1-$\sigma$ errors on that parameter (found by simply holding M/L and $\beta_o$ constant and varying . Most dwarfs have very little sensitivity to the rate at which the anisotropy decreases with radius, with the only exception being Draco. Many of the dwarfs (Leo I, Leo II, UMi and Sextans) are quite comfortable with constant anistropies, however, none of the dwarfs are consistent with an anisotropy that becomes highly biased towards tangential orbits in small increases of radius. It is difficult to be quantitative about how we could rule out MOND with any of these velocity anisotropies, since we use a very simple model for this and anisotropy could vary in a non-linear way, or there could be distinct stellar populations with completely different anisotropies. However, if instead of becoming biased towards tangential orbits at large radius, they were in reality biased towards highly radial orbits, there would be a serious problem because no M/L without a dark halo could fit the shape of the profiles. An example of this is given for the Sculptor dwarf in Fig \ref{fig:sculptor} where the red line shows the fit with a constant $\beta(r)=0.8$. No scaling with M/L can produce a good fit and it shows how precise knowledge of the velocity anisotropy would be extremely constraining to MOND, whereas any CDM halo can never be constrained to such detail and thus can never be falsified. 

\section{Results}

\begin{table*}
\centering
\tiny
\begin{tabular}[p]{|c|c|c|c|c|c|c|c|c|c|c|c|c|c|c|c} 
Name & $R_{MW}$ & c & $R_t$ & $M_v$ & $L_v$       & $V_r$ & $r_{\alpha}$& $\alpha_o$& $r_{\beta}$& $\beta_o$& $M/L_v$&$M/L_N$&$\chi^2_{red}$&n&Fig.3 linetypes\\
     & kpc      &   & pc    &       & $10^5\lsun$ & $\kms$     &  pc         &  &pc    \\
Carina&$101\pm5$ & 0.51$\pm$0.08 &$581\pm86$&-9.3$\pm$0.2&4.4$\pm$1.1&8&     2000&24&  $150_{-130}^{+1050}$&0.6&$5.6_{-2.9}^{+5.2}$& 80  &0.72 & 21 &black solid\\
Draco&$93\pm6$ & 0.72$\pm$0.05 &$1225\pm80$&-9.0$\pm$0.3&3.3$\pm$1.1&-294&    1500&14& $120_{-85}^{+150}$&0.93&$43.9_{-19.3}^{+29.0}$& 270 &0.65&19&black dashed\\
Leo I&$257\pm8$ & 0.39$\pm$0.07 &$1002\pm50$&-12.4$\pm$0.2&76$\pm$7&178&     3700&22&  $490_{-400}^{+\infty}$&0.1&$0.7_{-0.3}^{+0.65}$& 10  &0.47&15&black dotted\\
Sextans&$95.5\pm3$ & 0.98$\pm$0.14 &$3445\pm1141$&-9.7$\pm$0.2&6.3$\pm$1.4&75&  1200&6&  $1250_{-600}^{+8500}$&0.0&$9.2_{-3.0}^{+5.3}$& 102  &1.2&20&black dot-dashed\\
Fornax&$138\pm8$ & 0.72$\pm$0.05 &$2078\pm177$&-13.2$\pm$0.2&158$\pm$16&-36&  3200&14& $770_{-450}^{+1500}$&0.3&$1.4_{-0.35}^{+0.45}$&  11  &1.4&43&red solid\\
Sculptor&$87\pm4$ & 1.12$\pm$0.12 &$1329\pm107$&-11.1$\pm$0.2&28$\pm$4&95&  1200&14&   $170_{-140}^{+600}$ &0.5&$3.7_{-1.4}^{+2.2}$&    36  &0.7&33&red dashed\\
Leo II&$233\pm15$ & 0.48$\pm$0.1 &$554\pm68$&-9.9$\pm$0.3&7.6$\pm$3&22&   2500&30&   $280_{-255}^{+\infty}$&0.56&$1.85_{-1.1}^{+2.0}$&  57  &1.1&14&red dotted\\
UMi&$76\pm4$ & 0.64$\pm$0.05 &$1977\pm104$&-10.3$\pm$0.4&11.0$\pm$4.8&-87&  6000&25&   $5000_{-4750}^{+\infty}$ &0.73&$5.8_{-3.6}^{+6.5}$&  440  &0.86 &23&red dot-dashed\\
\end{tabular}
\caption{All distance and magnitude data from Mateo (1998), Galactocentric radial velocities from Wilkinson \& Evans (1999) and structural parameters from Irwin \& Hatzidimitriou (1995) except Draco's $V_r$ which is from Hargreaves et al. (1996). $R_{MW}$ is actually $R_{\sun}$, but since  satellites have Galactic latitudes such that $90^o<l<270^o$, this will only underestimate $R_{MW}$. Recalculated distance (Bellazzini et al. 2002 for both) and structural values  of are used for Draco (Odenkirchen et al. 2001) and Ursa Minor (Palma et al. 2003) which lead to new absolute magnitudes. Sextans has an updated distance from Lee et al. (2003) which gives a 20\% increased luminosity.  Leo I's luminosity and structural parameters come from Sohn et al. (2007) which makes use of the new distance modulus from Bellazzini et al. (2004). Luminosity for Leo II corrected for new distance from Bellazzini et al. (2005), as is Sculptor's from Piatek et al. (2006). The solar absolute magnitude is $M_v=4.8$. $W_o$ parameters for the King model are 1.97, 3.14, 1.51, 4.59, 3.14, 5.29, 1.85, 2.7. The Newtonian M/L come from dividing the dark mass at $r_{max}$ from Walker et al. (2007) by the V-band luminosity (except UMi which comes from Kleyna et al. 2002). n is the number of data points and $n_{d.f}$ is the number of degrees of freedom in the model (always 3).}
\end{table*}

The fits to the los velocity dispersion profiles of all 8 dwarfs are shown in Fig 1. The solid line is the best fit and the M/L is given in table 1 along with all other relevant data for each dwarf. The dashed lines are the 1-$\sigma$ errors in M/L. The reduced $\chi^2/(n-n_{d.f}-1)$, where n is the number of data points and $n_{d.f}$ is the number of degrees of freedom in the model (always 3), is very reasonable for each dwarf, even though the data are inherently noisy. A good example of this is the Fornax dwarf in which there are a larger number of data points (15) at large radii (from 500-1000pc) for which the individual errors are relatively small compared to the spread in the data. No straight (or even with curvature in one direction) line through these data points could obtain a low reduced $\chi^2$ and the line of best fit to these data is a very good one. 

The anisotropy profiles for each dwarf are plotted in Fig 2 against $R/R_c$ and the linetype is given in table 1. It is interesting that, apart from Ursa Minor and Draco, the other 6 dwarfs have very similar anisotropies going from isotropic or slightly radially biased orbits at their centres to more tangentially biased orbits at the edges.

To be comfortably explained in MOND, these dwarfs should have M/L of around 3 solar units. It is good to see that all four dwarfs associated with the Fornax-Leo-Sculptor stream have normal M/L. This is absolutely necessary because Leo I and Fornax are very massive compared to the rest which means the external gravity of the MW should not disturb them. Furthermore, the two Leo dwarfs are beyond 200kpc meaning even if they had a near pericentre to the MW, they should have returned to equilibrium by now.

The Carina and Ursa Minor dwarfs have borderline reasonable M/L ($5.6_{-2.9}^{+5.2}$ and $5.8_{-3.6}^{+6.5}$ respectively). When the 1-$\sigma$ error in M/L is considered along with deviations from spherical symmetry, coupled with other uncertainties in luminosity, distance, structure and possible tidal effects, they may well be safely within the bounds of normal M/L.

Although initially thought to have a large mass discrepancy, reanalysis of Ursa Minor's structural parameters and distance by Bellazzini et al. (2002) showed it to be more distant and extended, therefore it has a significantly larger luminosity which has reduced the M/L.

\subsection{Problem cases}
\protect\label{sec:pc}

There can be little doubt that Draco and now Sextans are both concerns for MOND. Their M/L's are uncomfortably high and we seek to find an explanation for this given that the other 6 are more or less compatible. First of all, it is troubling that these two have the noisiest velocity dispersions especially compared to the almost constant  velocity dispersions of Fornax, Sculptor, Leo I and even Carina. The dispersions seem to have regions where they are abnormally high and low, in addition to the presense of substructure in Sextans (Kleyna et al. 2004). Nevertheless, even if we only paid attention to the low values (i.e. with the lower dashed lines in Fig 1, the M/L are still unsatisfactory. Draco's luminosity and distance have recently been reviewed by Bellazzini et al. (2002), but Sextans still has a very poorly constrained structure and it would be of great benefit to have its King model parameters and integrated magnitude reanalysed. Still, it is worth bearing in mind that although the M/L are high, they are also very poorly constrained and they are only $\sim$2-$\sigma$ from acceptable values.

\subsection{The double impact of Galactocentric distance in MOND}
For a satellite like any one of the dwarfs studied here, the only influence the Newtonian gravity of the MW exerts is to carve out its orbital path. Any tidal forces are negligible and there is absolutely no difference to the internal gravity. In stark contrast, the MW's MOND gravity has a huge impact on the internal gravity of the low luminosity, nearby dwarfs: Carina, Sextans, Draco and Ursa Minor. As mentioned above, the external field effect renormalises the effective Newtonian Gravitational constant $G_{eff}= G/\mu({170^2 \over a_o R_{MW}})$ (as long as the external gravity is higher than the internal gravity). As $R_{MW}$ increases, $G_{eff}$ follows suit reducing all inferred M/L by the same factor. The key thing here is that if new observations increase the distance to a dwarf by 11\% as Lee et al. (2003) has done for Sextans (and Bellazzini et al. 2002, 2004, 2005 have done for other dwarfs) not only does the luminosity increase by 21\%, which obviously drops any M/L by 17\%, but the reduction in the external field effect allows the M/L to drop a further 10\% (in that particular case). So, the increased distance of less than 10kpc for Sextans dropped the M/L from 15 to almost 9.

\subsection{Sextans: poorly constrained light profile}
Sextans has a M/L=$9.2_{-3.0}^{+5.3}$ which is impossible from population synthesis point of view. However, this M/L assumes we know the luminosity of Sextans well. In fact, Sextans is the lowest surface brightness dwarf we study here and it's light profile has not been examined in detail since before Mateo (1998). All of the other dwarfs that have had their luminosity re-evaluated have had it pushed up far beyond the original 1-$\sigma$ error: UMi, Draco, Carina, Leo I and Leo II. In fact, the first four of those five mentioned saw their luminosities double. When Sextans has its luminosity re-integrated with new photometry, we expect it to rise by at least 50\%, but not necessarily by a factor of 3 which would put it in the acceptable zone. This will probably leave it with a similar M/L as Carina and UMi, requiring a modest amount of tidal heating (see next section) to be consistent with the los velocity dispersions.

At the moment however, it should be noted that although the M/L of Carina and UMi are higher than 3, uncertainties in their los velocity dispersions as well as their luminosity and distance mean they are in agreement within the observational errors. Still, the fact that they are all coherently, higher than preferred compells us to consider why in the next section.

\subsection{Impulsive tidal heating}

The most enticing resolution of the three moderately high M/L dwarfs are that Sextans, Carina and UMi have recently had a close pericentre of the MW and have been slightly tidally heated. The crossing times (2R/$\sigma_{los}$) at the last measured los dispersion radius are 120 (Carina), 300 (Sextans), 150 (Umi) and 110 (Draco) Myr, but recall that relaxation times in MOND are higher than in Newtonian gravity as calculated from galaxy mergers by Nipoti et al. (2007) and Tiret \& Combes (2007). Using a simple back stepping orbit integrator taking the current Galactocentric radii and radial velocities, with gravity calculated from a flat rotation curve with circular speed of 170$\kms$ (see Xue et al. 2008), we found it has been roughly 700 (Carina), 500 (Sextans) and 850 (UMi) Myr since the satellites had their last pericentre. Draco's apocentre is over 400kpc, which means it is probably on its first passage through the MW since this orbit integration neglects the external field of M31 and large scale structure on the MW and despite this it is 5600Myr since Draco left. 

There is something key to remember here in which MOND differs starkly from Newtonian gravity. A low surface brightness dwarf galaxy has dynamical stability imparted by the MOND enhanced gravity, much like the DM halo provides for Newtonian gravity. If we make the crude but fair assumption, for clarity's sake, that the external field of the MW dominates the internal accelerations of the 3 satellites, then the MOND effect is to renormalise Newton's constant upwards which enhances the internal gravity. So $G_{eff} = G/\mu(g_{ex}/a_o)\sim G a_o/g_{ex}$. 

As the satellites approach the MW this stability is lost due to the rising external field which bleeds the internal gravity of the dwarf. This of course doesn't happen in Newtonian gravity and the dwarfs can withstand tidal forces more easily (the DM halo stays intact). This makes it far easier to perturb a MOND dwarf. The radii below which tidal effects begin to affect the dwarfs $\sim$40kpc  as opposed to $<$10kpc for Newtonian gravity with the observed DM halos (see Brada \& Milgrom 2000b). 

However, currently, none of the dwarfs studied here are strictly within the range of radii with measured dispersions, in the non-adiabatic/impulsive regime. However, they may have just left this region. We re-emphasise their point that tidal effects begin to impose themselves at much larger Galactocentric radii in MOND. We slightly repackage this idea by looking at the rate of change of the external field strength during one crossing time at the core radius of each dwarf i.e. ${d \over dt}\left({g_{ex}\sigma_{los} \over R_c}\right) \sim {170^2 \over R_{MW}}-{170^2 \over R_{MW}-2V_rR_c/\sigma_{los}}$ plotted against M/L for each dwarf in Fig \ref{fig:dsph}). 

From Fig 3 one can see that low Galactocentric radius, luminosity and surface brightness dwarfs have higher M/L. This is unsurprising because these properties make the dwarf more susceptable to tidal effects. High rate of change of external field is where Draco stands out with its enormous M/L. Although the dwarfs are not in the non-adiabatic regime, the magnitude of the external field will have been varying rapidly in the past. It is unknown at present how serious an effect this is, but it needs to be studied with realistic, high resolution MOND simulations (see Nipoti et al. 2007; Tiret \& Combes 2007) for dwarf galaxies in the external field of the MW, but also for spirals passing through the centres of clusters. This rapidly varying external field we predict is destroying the integrity of spirals in clusters (see Rubin et al. 1999) and also imposes a morphology-position relation in clusters more easily than galaxy harassment (Moore 1999).

Whether Sextans or any of the others have recently had a close pericentre passage will only be known after accurate proper motions are published. The proper motions of Ursa Minor are known well enough only to say the pericentre lies between 10kpc and 76kpc (Piatek et al. 2005) at 1-$\sigma$. Brada \& Milgrom (2000) showed that before tidal effects are felt there is a region where non-adiabatic (impulsive) effects due to the quickly changing external field heat up the dwarf without breaching it. There may be insufficient time for the satellite to regain equilibrium after leaving this zone, especially since they are still external field dominated.


Of course, when accurate proper motions become available, if the opposite is true and they are currently at pericentre and on a reasonably circular orbit, they become quite a serious problem for MOND. Finally, as mentioned in \S\ref{sec:pc}, Sextans should be reanalysed in more detail to tighten the errors on its light profile and magnitude.

\subsection{Draco: tidal disruption}

If indeed Sextans has been heated by a close pericentre, the biggest issue we are left with is Draco with M/L=$43.9_{-19.3}^{+29.0}$. Draco is now moving towards the MW at $\sim294\kms$ which means it is most likely on its first passage of the MW. S\'egall et al. (2007) showed that Draco has a symmetric, unperturbed density profile with no evidence for tidal tails - ``Draco has led a quiet existence". This is in line with the fact that it has been on a direct path towards the MW since its creation. However, if its tangential velocity is negligible, its tidal tails may line up along the los.

Since Draco cannot have been tidally heated by a near pericentre passage due to the timescales involved, this leaves us with only one conclusion i.e. that Draco is in the throes of tidal disruption, which is perfectly possible in MOND, but unimaginable in Newtonian gravity. At its apocentre, Draco would have experienced very little external field and thus its low surface density would be enough to maintain its dynamical integrity due to the MOND enhanced gravity. This would make its internal gravity at R=1000pc $\sim \sqrt{3.6\times4.4\times10^{-3}\times3\times3.3\times10^5/1000^2}\sim a_o/30$ (this assumes no external field). However, currently it is at 93kpc and thus the internal gravity is merely $4.4\times10^{-3} \times a_o {93\times 10^3 \over 170^2} \sim  a_o/70$, reduced because of the growing external field. This loss in binding energy might have recently thrown Draco into the early stages of tidal disruption. Mu\~noz et al. (2007) have showed with high resolution N-body simulations that even modest tidal disruption near the edges of dwarfs can have a dramatic impact on the observed dynamics. This again must be confirmed by numerical simulations of Draco on a plunging (radial) orbit from a large radius in MOND with a reasonable M/L and velocity dispersion found by solving the Jeans equation. Either Draco will begin to disrupt at $\sim 100kpc$ and its los velocity dispersions will appear artificially high, or it will have dispersions consistent with the Jeans equation at all radii down to at least 50kpc. 

\subsection{Dwarf fly-bys}

Is it possible that the dwarfs are co-responsible for their own large dispersions? Draco and Ursa Minor are separated by approximately 20kpc, so the largest external gravity that UMi could inflict on Draco is a mere $Ga_o/g_{ex}(M/L_v)L_{sat}d^{-2} \sim 4.4\times10^{-3} \times 10 \times 3 \times 1.1 \times 10^6 \times(20000)^{-2} \sim a_o/10000$, and since they are moving towards each other now, they were never any closer in the past, so there is no chance of this being important. If say the Fornax satellite conspired to collide with Sextans it would have to pass within $d\sim \sqrt{G{a_o \over g_{in}g_{ex}}(M/L_v)L_{sat}} \sim \sqrt{4.4\times10^{-3}}\sim 4500pc$. That would balance the internal and external accelerations because at 1kpc the internal gravity of Sextans is $a_o/20$. This is at least possible given that they are currently moving away from each other. Accurate proper motions would resolve this.

\subsection{Tidal tails and ellipticity}
It was noted by Klimentowski et al. (2007) and Lokas et al. (2008), tidal tails along the line of sight can have an adverse effect on the los velocity dispersion. They showed using high resolution N-body simulations as comparisons that the velocity dispersion is boosted by the interlopers, giving the appearance of a high M/L. This may indeed reduce the M/L for several of the dwarf spheroidals studied here, but it remains to be seen whether the effect is significant enough to reduce Draco's M/L by an order of magnitude.

Linked to this is the idea that deviations from spherical symmetry could produce the high dispersions. As above, if the major axis was pointing along the los, then this would have the same effect as tidal tails. So significant help could be granted by tidal streams or deviations from spherical symmetry.

\subsection{Relativistic MOND}
The field of relativistic MOND kicked off by Bekenstein (2004) is growing fast and the best bet at the moment is the Einstein Aether theory (see Zlosnik, Ferriera \& Starkman 2007), however, one should see Milgrom (2008) for a review of possible fundamental origins for MOND. Currently, it is unknown exactly how systems, like dwarfs, with rapidly time varying external gravity might fare in these theories, but it may be the case that the vector-field boosts the internal gravity to a greater extent than simple MOND.

\subsection{Dust}

The easiest solution to the problem would of course be if there was simply a large dust sheet obscuring most of the light of the three galaxies. In a recent paper by Driver et al. (2007) they postulated that more than 90\% of the light in galaxy bulges (dwarfs are many orders of magnitude removed bulges in most properties) can be obscured in the V-band by dust.  This effect actually is still not enough and the remaining light would have to be attenuated by dust in the Milky Way. Draco and Ursa Minor are on the side of the Milky way opposite the bulge from the solar position, so it is not even as if we are looking through high stellar, dust and gas density towards them. Finally, since deep Color-Magnitude-Diagrams exist for all these galaxies, if they were affected by significantly larger extinction than presently believed it would have been noted long ago.

\subsection{Trends with key variables}
In Fig \ref{fig:trends} we plot the M/L against four variables which describe the dwarfs in order to elucidate trends, although we are clearly dealing with large errors and small number statistics. In the top left panel, we plot M/L against Galactocentric distance. There is an obvious division between Fornax and the two Leo dwarfs beyond 130kpc and the other 5 less than 100kpc from the centre of the MW. In addition, there is some evidence for a trend shown in the top right panel for lower surface brightness dwarfs to have higher M/L as well as those with an external field that is varying more rapidly.

However, none of those three independent variables cut a tight correlation with M/L because none of them are the cause of the high M/L, they are merely strong indicators. The smaller the Galactocentric radius, the higher the external field and the more energy there is available to perturb the dwarf. The lower the luminosity, the more susceptible the dwarf is to being perturbed by the external field. The radial velocity (bottom left) is a poor indicator because any perturbing of the dwarfs was done near pericentre. This again shows why Galactocentric radius is just an indicator because it simply means the dwarf was more likely to have been at smaller radii but for all we know it may currently be at pericentre (Bellazzini et al. 1996). It is not surprising that Draco with its large M/L has a low Galactocentric radius, surface brightness and large velocity towards the MW as it may well be tidally disrupting.

\begin{figure}
\includegraphics[angle=0,width=8.0cm]{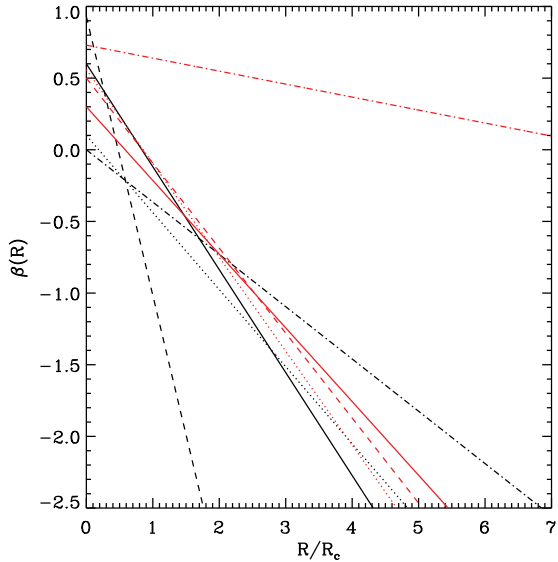}
\caption{Anisotropy profiles, $\beta$(r), for all 8 dwarfs. See table 1 for the corresponding linetypes.}
\end{figure}

\begin{figure*}
\def\subfigtopskip{0pt} 
\def\subfigbottomskip{4pt}
\def\subfigcapskip{1pt}
\centering

\begin{tabular}{cc}

\subfigure{\label{fig:mlrmw}
\includegraphics[angle=0,width=8.0cm]{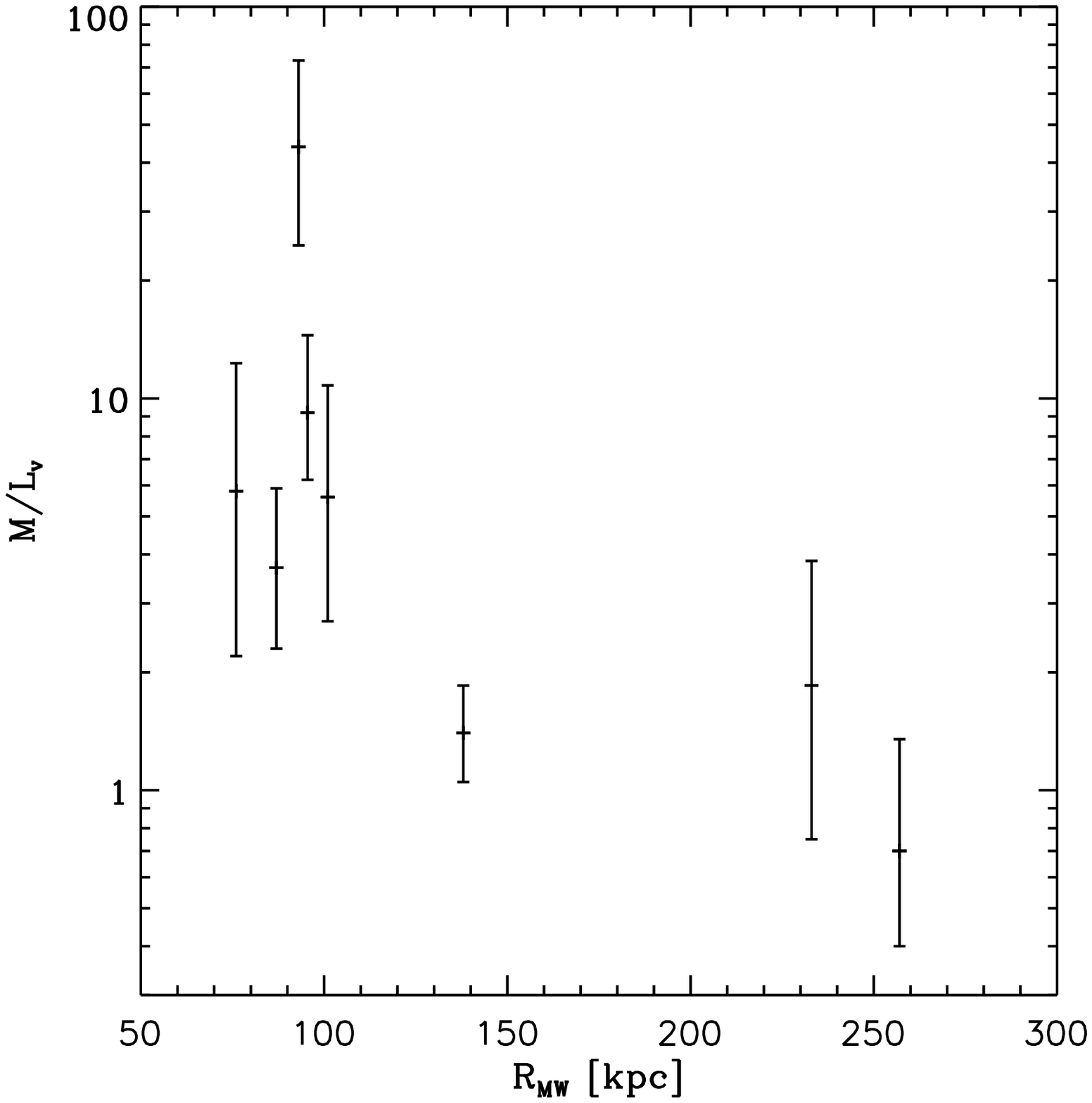}
}
&
\subfigure{\label{fig:mlgex}
\includegraphics[angle=0,width=8.0cm]{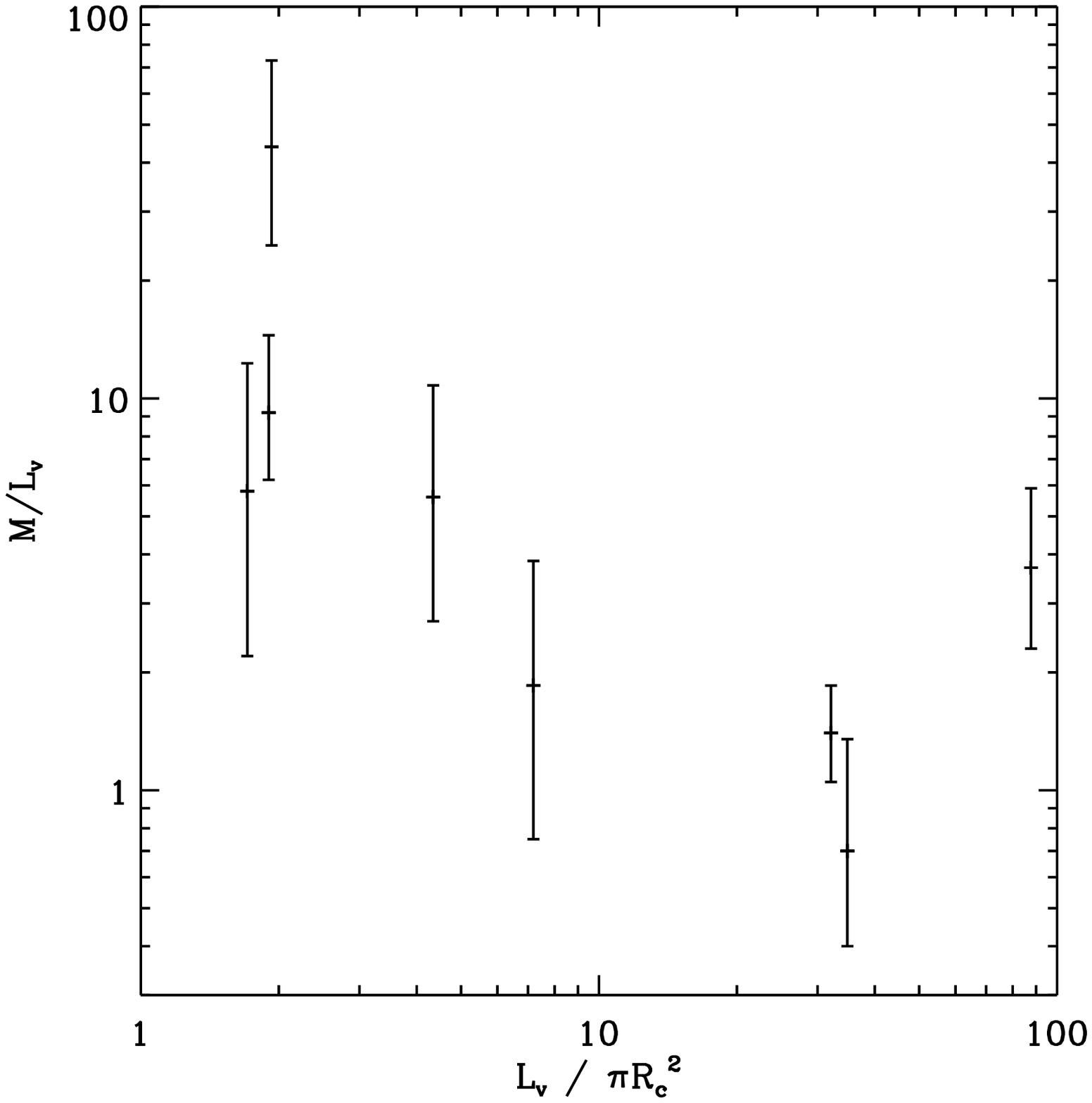}
}
\\
\subfigure{\label{fig:mlvr}
\includegraphics[angle=0,width=8.0cm]{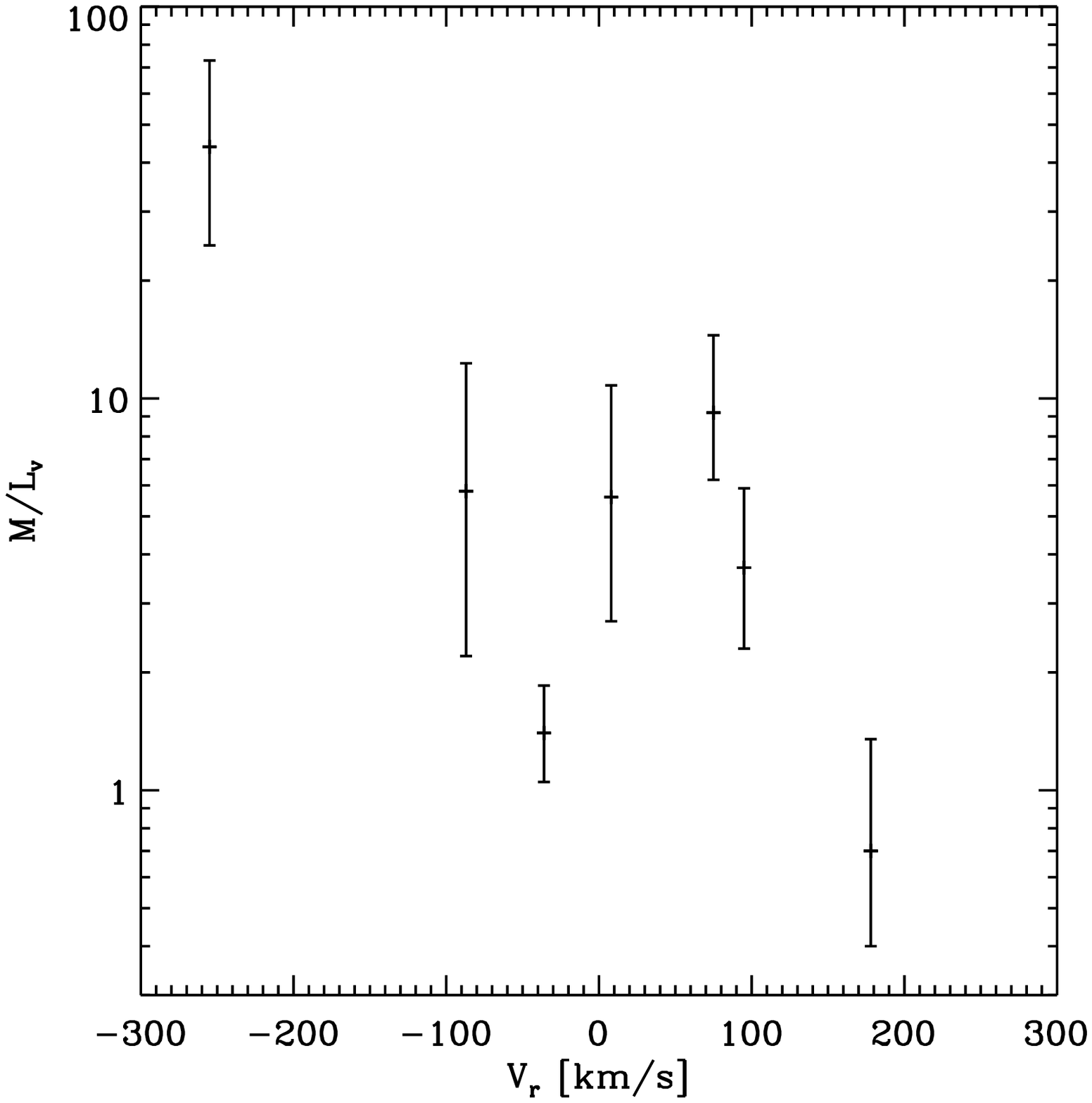}
}&
\subfigure{\label{fig:mldtgex}
\includegraphics[angle=0,width=8.0cm]{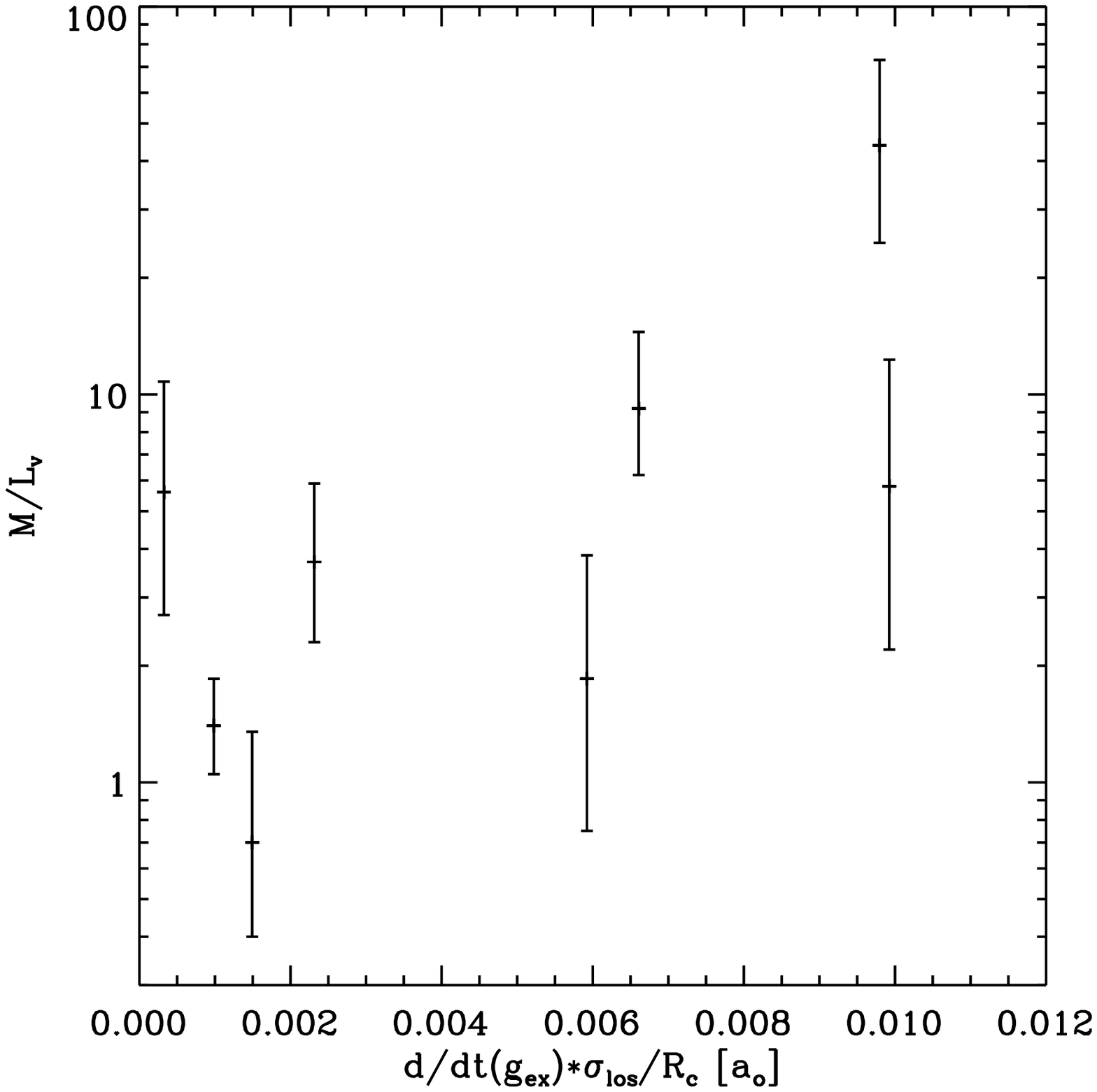}
}
\\
\end{tabular}

\caption{Shows M/L against several variables for each dSph. (a) Galactocentric radius, which implies external field strength (b) V band surface density (c) Galactocentric radial velocity (d) time derivative of the external field in units of the dSph's dynamical time at the core radius. Note the three dwarfs with large time derivative of the external field are Draco, Ursa Minor and Sextans.}
\label{fig:trends}
\end{figure*}

\section{The Magellanic Stream}

The new proper motions of the Magellanic Clouds (Kallivayalil et al. 2006ab) show conclusively that the Large Magellanic Cloud (LMC) and Small Magellanic Cloud (SMC) binary are currently at pericentre (Besla et al. 2007, hereafter B07) in their orbit of the MW. The creation of the Magellanic stream, a highly collimated gas cloud arcing $180^o$ across the sky and emanating from the cloud's position. Several puzzling features plague the stream such as the fact the stream does not trace the orbit of the LMC in any favourable CDM (B07, Fig. 9) or MOND orbit (See X.F. Wu et al. in prep), there are very few stars in the cloud (B07) and at a pericentre of $>50kpc$ the stream cannot be tidally induced. Eyeballing Fig. 18 of B07 one can see the motion of the SMC is significantly influenced by the LMC allowing its orbit to trace the stream. As to how the gas was stripped from the SMC, we need only consider the ram pressure of the gas in the LMC. As mentioned above, on close approach to the LMC, the SMC's internal gravity is damped by both by the MW and more importantly, by the LMC.

The criteria for ram pressure stripping (Roediger \& Bruggen 2007) is for the gravitational restoring force per unit area,
\beq
\protect\label{eqn:fpa}
f_{grav}(R)=\Sigma_{g,SMC}(R)g_z(R),
\eeq
to be less than the ram pressure
\beq
\protect\label{eqn:ram}
P_{ram}=\rho_{g,LMC} V_{rel}^2.
\eeq
where $\Sigma_{gas}$ is the surface density of gas in the SMC, $g_z$ is the SMC's gravity in the direction of the collision, $\rho_g$ is the density of gas in the LMC and $V_{rel}$ is the relative velocity between the LMC and SMC at collision.

The relative velocity between the pair at the moment is around $V_{rel}=100\kms$. The total gas mass of the LMC is $M_{g,LMC}=5\times10^8\msun$, so if it has dimensions (R,z)=(8.5kpc, $\pm$0.2kpc) then the average density of gas is $\rho_{g,LMC}={5\times10^8\msun \over \pi (8.5\times10^3)^2\times400}\sim0.02\msun pc^{-3}$. The mass of gas in the SMC is $M_{g,SMC}=4.2\times10^8\msun$ (Stanimirovi\'c, Staveley-Smith and Jones 2004)so for a typical radius of 2.5kpc, the surface density of gas is $\Sigma_{g,SMC}=22\msun pc^{-2}$.

The key input to this equality and the only variable that changes from MOND to Newtonian dynamics is the gravity of the SMC in the direction of the collision. The total baryonic mass of the SMC is around $M_{SMC}\sim8.5\times10^8\msun$ (Stanimirovi\'c et al. 2004). With the LMC so close to it, the SMC's internal gravity is simply Newtonian with a renormalised gravitational constant, $G_{eff}=G a_o/g_{LMC}$. At 3.5kpc, the rotation curve of the LMC is flat (Piatek, Pryor and Olszewski 2007) at 120$\kms$. At this radius, $g_{LMC} \sim 1.2a_o$. This means that $G_{eff}<2G$, whereas if the LMC was not there, and (more or less) compared to Newtonian gravity, $G_{eff}\sim7G$.

With these special case parameters, the gravity must be $g_z \sim 2.5a_o$ to retain the gas, whereas the gravity is less than this: $g_{z,M}(R=3.5kpc)\sim a_o/6$ and the Newtonian value is 3-4 times larger. This shows that ram pressure stripping in MOND is easier which is completely contrary to the naive assumption ignoring the external field effect. This can easily be the mechanism which removes the gas from the SMC and creates the Magellanic Stream. This has further implications for the ram pressure stripping of normal spirals in clusters and other satellite galaxies of the MW.

\section{Conclusion}
Here I have studied the los velocity dispersion profiles of 8 of the MW's dwarf satellites in MOND. We have shown that 6 of the dwarfs (Leo I, Leo II, Fornax, Sculptor, Ursa Minor and Carina) have acceptable M/L within their errors. Despite Carina and Ursa Minor having M/Ls of $5.6_{-2.9}^{+5.2}$ and $5.8_{-3.6}^{+6.5}$ respectively, it is pointless to be too concerned about them because they have M/L of 3 within 1-$\sigma$ and uncertainties in luminosity, ellipticity, structure and equilibrium could easily account for this discrepancy. On the other hand, the remaining 2 (Draco and Sextans) have M/L completely inconsistent with the stellar populations. The possible causes of this have been discussed in depth and 3 testable effects that cause the discrepancies have been suggested.

First of all, the photometry of Sextans is extremely old (more than 10 years) and so I suggest K-band images of the dwarf should be acquired to alleviate any doubts on the luminosity of the dwarf and how the light is distributed. However, this is likely to leave Sextans with a M/L similar to Carina, UMi and Sextans ($\sim 5$). To be certain of the effects of non-adiabatic tidal heating of the dwarfs, we need accurate proper motions to make sure Sextans is currently not at pericentre and that Draco is on a highly elongated orbit plunging towards the MW.

If this is the case, then realistic, high resolution simulations (like those of Nipoti et al. 2007 and Tiret \& Combes 2007) must be run in order to confirm that the non-adiabatically changing external field effect can indeed make the small boost to the los velocity dispersions in these satellites over the given orbits and timescales and even tidally disrupt Draco. If it turns out that the external field effect is unable to disrupt Draco given its orbital history then it poses a significant problem for MOND.

It is quite remarkable that using only the stellar mass of the dwarfs, we can successfully match the los velocity dispersion profiles with sensible falling anisotropies and get excellent reduced $\chi^2$ for all 8 dwarfs and 6/8 have good M/L. As for draco, the unknown orbital history is vital to the comprehension of its large M/L and until it is known, we cannot confirm the dwarf galaxies of the MW as a success or failure of MOND.

\section{acknowledgements}
GWA thanks Michele Bellazzini, Ricardo Mu\~noz, Alan McConnachie, Mario Mateo, Moti Milgrom, Stacy McGaugh and Benoit Famaey for discussions. This research was supported by an STFC fellowship.

\label{lastpage}


\begin{thebibliography}{}
\bibitem{afz}Angus G.W., Famaey B., Zhao H.S., 2006, MNRAS, 371, 138
\bibitem{aszf}Angus G.W., Shan H.Y., Zhao H.S., Famaey B., 2007a, ApJ, 654, L13
\bibitem{aftcz}Angus G.W., Famaey B., Tiret O., Combes F., Zhao H.S., 2008, MNRAS, 383, L1
\bibitem{afb}Angus G.W., Famaey B., Buote D., 2007b, preprint(arXiv:0709.0108)
\bibitem{ab}Angus G.W., McGaugh S.S., 2008a, MNRAS, 383, 417
\bibitem{ab}Angus G.W., McGaugh S.S., 2008b, submitted MNRAS
\bibitem{BM84} Bekenstein J., Milgrom M., 1984, ApJ, 286, 7
\bibitem{asq}Bekenstein J.D., 2004, PhRvD, 70, 083509
\bibitem{bekrev} Bekenstein J.D., 2006, Contemporary Physics, 47, 387
\bibitem{asq}Bellazzini M., Ferraro F.R., Origlia L., Pancino E., Monaco L., Olivia E., 2002, AJ, 124, 3222
\bibitem{asq}Bellazzini M., Gennari N., Ferraro F.R., Sollima A., 2004, MNRAS, 354, 708
\bibitem{asq}Bellazzini M., Gennari N., Ferraro F.R., 2005, MNRAS, 360, 185
\bibitem{bes}Besla G., et al., 2007, ApJ, 286, 7
\bibitem{bourn} Bournaud F., et al., 2007, Science, 316, 1166
\bibitem{asq}Brada R., Milgrom M., 2000a, ApJ, 531, 21
\bibitem{asq}Brada R., Milgrom M., 2000b, ApJ, 541, 556
\bibitem{asq}Ciotti L., Binney J.J., 2004, 351, 285
\bibitem{asq}Driver S.P. et al., 2007, MNRAS, 379, 1022
\bibitem{GFC}Gentile G., Famaey B., Combes F., Kroupa P., Zhao, H.S., Tiret O., 2007, A \& A, 472, L25
\bibitem{Gil}Gilmore G., Wilkinson M., Kleyna J., Koch A., Evans W., Wyse R.F.G., Grebel E.K., 2007, ApJ, 663, 948
\bibitem{Gil}Gilmore G., Wilkinson M., Kleyna J., Koch A., Evans W., Wyse R.F.G., Grebel E.K., 2007, Nuclear Physics B (Proc. Suppl.), 173, 15
\bibitem{asq}Hargreaves J.C., Gilmore G., Irwin M.J., Carter D., 1996, MNRAS, 282, 305
\bibitem{asq}Ibata R.A., Gilmore G., Irwin M.J., 1994, 370, 194
\bibitem{ih}Irwin M., Hatzidimitriou D., 1995, 277, 1354
\bibitem{jac}Kallivayalil N., et al., 2006a, ApJ, 638, 772
\bibitem{jac}Kallivayalil N., et al., 2006b, ApJ, 652, 1213
\bibitem{asq}Kleyna J.T., Wilkinson M.I.,, Evans N.W., Gilmore G., 2004, MNRAS 354, L66
\bibitem{asq}Klimentowski J., Lokas E.L., Kazantzidis S., Prada F., Mayer L., Mamon G.A., 2007, MNRAS 378, 353
\bibitem{koch}Koch A., Kleyna J.T., Wilkinson M.I., Grebel E.K., Gilmore G.F., Evans N.W., Wyse R.F.G., Harbeck D.R., 2007, AJ, 134, 556
\bibitem{lee}Lee M.G., et al., 2003, AJ, 126, 2840
\bibitem{lok01}Lokas E., 2001, MNRAS, 327, L21
\bibitem{asq}Lokas E., Mamon G.A., Prada F., 2005, preprint (arXiv:astro-ph/0508668v2)
\bibitem{lok08}Lokas E., Klimentowski J., Kazantzidis S., Mayer L. 2008, preprint (arXiv:0804.0204)
\bibitem{asq}Lynden-Bell D., 1983, E. Asthanassoula (ed.), Internal Kinematics and Dynamics of Galaxies, 89-92
\bibitem{McG5} McGaugh S.S., 2004, ApJ, 609, 652
\bibitem{McG5} McGaugh S.S., 2005, ApJ, 632, 859
\bibitem{MdB} McGaugh S.S., de Blok W.J.G., 1998, ApJ, 499
\bibitem{MdB} McGaugh S.S., Schombert J. M., Bothun G. D., de Blok W.J.G., 2000, ApJ, 533, L99
\bibitem{asq}Mateo M., 1998, ARAA, 36, 435
\bibitem{asq}Metz M., Kroupa P., 2007, preprint(arXiv:astro-ph/0701289)
\bibitem{Mila} Milgrom M., 1983, ApJ, 270, 365
\bibitem{Mila} Milgrom M., 1995, ApJ, 455, 439
\bibitem{tdgm} Milgrom M., 2007a, ApJ, 667, L45
\bibitem{tdgm} Milgrom M., 2007b, preprint (arXiv:0712.4203)
\bibitem{tdgm} Milgrom M., 2008, preprint (arXiv:0801.3133)
\bibitem{asq}Milgrom M., Sanders R.H., 2007, preprint (arXiv:0709.2561v1)
\bibitem{moore99} Moore B., Lake G., Quinn T., Stadel J., 1999, MNRAS, 304, 465
\bibitem{asq}Mu\~noz R.R., et al., 2005, ApJ, 631, L137
\bibitem{asq}Mu\~noz R.R., et al., 2006, ApJ, 649, 201
\bibitem{mun2}Mu\~noz R.R., Majewski S.R., Johnston K.V., 2007, preprint(arXiv:0712.4312)
\bibitem{pal} Odenkirchen M., et al., 2001, AJ, 122, 2538
\bibitem{pal}Palma C., Majewski S.R.,Siegel M.H., Patterson R.J., Ostheimer J.C.,Link R., 2003, AJ, 125, 1352
\bibitem{asq}Piatek S., et al., 2005, AJ, 130, 95
\bibitem{asq}Piatek S., et al., 2006, AJ, 131, 1445
\bibitem{asq}Piatek S., Pryor C., Olszewski E.W., 2007,preprint (arXiv:0712.1764)
\bibitem{asq}Roedriger E., Bruggen M., 2007, MNRAS, 380, 1399
\bibitem{asq}Rubin V.C., Waterman A.H., Kenney J.D.P., 1999, ApJ, 118, 236
\bibitem{ARAA} Sanders R.H., McGaugh S.S., 2002, ARAA, 40, 263
\bibitem{asq}S\'egall M., Ibata R.A., Irwin M.J., Martin N.F., Chapman S., 2007, MNRAS, 375, 831
\bibitem{asq}Sohn S.T., et al., 2007, ApJ, 663, 960
\bibitem{asq}Sollima A., 2004, MNRAS, 354, 708
\bibitem{asq}Stanimirovi\'c S., Staveley-Smith L., Jones P.A., 2004, ApJ, 604, 176
\bibitem{asq}Tiret O., Combes F., 2007, preprint(arXiv:0712.1459)
\bibitem{wu} Wu X., Zhao H.S., Famaey B., Gentile G., Tiret O., Combes F., Angus G.W., Robin A.C., 2007, ApJ, 665, L101
\bibitem{Xue} Xue X.-X., et al., 2008, preprint(arXiv:0801.1232)
\bibitem{asq}Zlosnik T., Ferriera P.G., Starkman G.D., 2007, preprint (arXiv:0711.0520)
\end{thebibliography}
\end{document}